\documentclass[aps,nofootinbib]{revtex4}
\usepackage{stmaryrd}
\usepackage{epsfig}
\usepackage{amsfonts}
\usepackage{amsmath}
\usepackage{graphicx}
\usepackage{color}
\usepackage{amssymb,amsmath,psfrag,slashed,graphicx}

\def \t {\widetilde}

\allowdisplaybreaks[4]
\newif\ifContLineOne
\newif\ifContLineTwo
\newif\ifContLineThree

\begin{document}

\title{Light-cone distribution amplitudes of vector meson in\\ large momentum effective theory}

\author{Ji Xu$^1$~\footnote{xuji1991@sjtu.edu.cn}, Qi-An Zhang$^2$~\footnote{zhangqa@ihep.ac.cn } and Shuai Zhao$^1$~\footnote{shuai.zhao@sjtu.edu.cn} }
\affiliation{  $^1$INPAC, Shanghai Key Laboratory for Particle Physics
and Cosmology,\\
MOE Key Laboratory for Particle Physics, Astrophysics and Cosmology,\\
 School of Physics and Astronomy, Shanghai Jiao Tong
University, Shanghai, 200240,   China\\
$^2$Institute of High Energy Physics, Chinese Academy of Sciences, Beijing 100049, China,\\
School of Physics, University of Chinese Academy of Sciences, Beijing 100049, China}

\begin{abstract}

We investigate the leading twist light-cone distribution amplitudes (LCDAs) of vector meson in the framework of large momentum effective theory.
We derive the matching equation for the LCDAs and quasi distribution amplitudes.
The matching coefficients are determined to one loop accuracy, both in the ultraviolet cut-off and dimensional regularization schemes.
This calculation provides the possibility of studying the full $x$ behavior of LCDAs and extracting LCDAs of vector mesons from lattice simulations.

\end{abstract}

\maketitle

\section{Introduction}

The light-cone distribution amplitudes (LCDAs),  defined by the matrix elements of light-cone separated field operators, are essential for the studying of exclusive processes and hadron structures. They describe the probability amplitudes of finding Fock states in a hadron.
The amplitude of an exclusive process can be factorized as the convolution of LCDA and hard kernel, if the collinear factorization theorem is established~\cite{Lepage:1980fj}. The LCDAs of vector mesons are of particular interest since they are necessary in the theoretical analysis of exclusive processes involving vector mesons, for example, the production of vector mesons at high energy colliders, and $B$ meson decays like $B\to V$, where $V=\rho, K^*, \phi$, etc. Among these processes, the $B\to K^*\ell^+\ell^-$ and $B_s\to \phi \ell^+\ell^-$ are considered to be sensitive to the new physics. Measurement of the angular distributions and the lepton flavor universality violation may shed light on the evidence of new physics~\cite{Aaij:2014ora,Aaij:2015oid,Aaij:2017vbb}.
The determination of LCDAs is thus also important for the searching of new physics beyond the Standard Model. Furthermore, LCDAs are also crucial in light-cone QCD sum rules (LCSR), perturbative QCD approach (PQCD) and so on.
 Because of their non-perturbative nature, LCDAs can not be calculated with QCD perturbation theory.

The LCDAs of mesons have been studied intensively due to their significance in phenomenology. Among the mesons, the LCDAs for pseudoscalar mesons like pion
and kaon are relatively well understood, since their LCDAs are much simplier. The LCDAs for vector mesons, e.g., the $\rho$ meson, are more complex since the vector meson can be either longitudinally or transversely polarized.
The LCDAs of vector mesons have been studied in various approaches, e.g, the QCD sum rules~\cite{Chernyak:1983ej,Ball:1996tb,Fu:2016yzx}, the lattice QCD (LQCD) calculation, etc.
 However, at present the LCDAs of $\rho$ meson are only calculated up to the second moment in LQCD approach~\cite{Arthur:2010xf,Braun:2016wnx}.

A novel strategy of evaluating light-cone correlators is the large momentum effective theory (LaMET), in which the full $x$ ($x$ is the longitudinal momentum fraction) dependence can be calculated~\cite{Ji:2013dva,Ji:2014gla}. In LaMET, instead of calculating light-cone correlation matrix elements, one can first evaluate the corresponding equal-time corralators, which can be simulated on the lattice.
The matrix elements defined by these equal-time correlators are the so called quasi quantities, e.g.,
quasi parton distribution functions (quasi-PDFs), quasi distribution amplitudes (quasi-DAs), etc. The quasi and light cone quantities have the same infrared (IR) structure, but their ultraviolet (UV) behaviors are different, the difference is involved in the matching coefficient.  Under the large $P_z$ limit, the quasi observables can be factorized as the convolution of perturbatively calculable coefficients and the standard light-cone observables.
 With such factorization formula, one can extract light-cone observables from lattice simulation.
Some other related proposals, e.g., the lattice cross section approach~\cite{Ma:2014jla,Ma:2017pxb} and the pseudo PDFs~\cite{Radyushkin:2016hsy,Radyushkin:2017cyf,Orginos:2017kos} are also developed. The LaMET has been studied to explore the quark PDFs~\cite{Xiong:2013bka}, gluon PDFs~\cite{Wang:2017qyg,Wang:2017eel}, transverse momentum dependent (TMD) PDFs~\cite{Ji:2014hxa,Ji:2018hvs}, generalized parton distributions (GPDs)~\cite{Ji:2015qla,Xiong:2015nua}, as well as pion and kaon's distribution amplitudes~\cite{Zhang:2017bzy,Bali:2017gfr,Chen:2017gck} and the LCDAs of heavy quarkonia~
\cite{Jia:2015pxx}. The renormalizability of quasi-PDF has been established recently~\cite{Ji:2015jwa,Constantinou:2017sej,Ji:2017oey,Ishikawa:2017faj}, and the non-perturbative renormalization schemes such as regularization invariant momentum subtraction (RI/MOM) have also been employed to study the quasi-PDFs~\cite{Alexandrou:2017huk,Constantinou:2017sej,Chen:2017mzz,Lin:2017ani,Stewart:2017tvs}.
LQCD calculations on quasi-PDFs shows the feasibility of evaluating PDFs and LCDAs from the first principle of QCD~\cite{Lin:2014zya,Alexandrou:2015rja,Chen:2016utp,Alexandrou:2016jqi,Zhang:2017bzy}.
Thus LaMET provides one more approach of accessing LCDAs of vector mesons by lattice simulation. Before the lattice evaluation is performed, it is necessary to determine the matching coefficient between the LCDAs and quasi-DAs, in QCD perturbation theory.

The present paper is devoted to the perturbative matching between the quasi and light-cone distribution amplitudes of vector mesons in LaMET.
We will study the twist-2 LCDAs of vector meson and the corresponding quasi-DAs.
The main aim of this work is to derive the matching equation for quasi and light cone distribution amplitudes.
To do this, we will calculate the one loop corrections to both the quasi and light-cone distribution amplitudes, then work out the matching coefficients to one loop accuracy. This work will provide the possibility of extracting LCDAs of vector meson from future lattice simulations.

The rest of this paper is organized as follows. In Sec.~II, we present the definitions of twist-2 LCDAs for the transversely and longitudinally polarized states,
and their corresponding quasi-DAs. In Sec.~III, we calculate the one-loop corrections to the LCDAs and quasi-DAs, in the UV cut-off scheme. In Sec.~IV, the LaMET matching equation will be derived.
 We summarize in Sec.~V. The results under dimensional regularization and matching coefficients with a finite UV cut-off $\Lambda$ will be arranged in the Appendices.

\section{Definitions of light-cone and quasi distribution amplitudes}

Before introducing the quasi-DAs, we first revisit the LCDAs.
We adopt the light-cone coordinate system to discuss the LCDAs. In light-cone coordinate system, any four-vector $a$ can be expressed as $a^{\mu}=(a^+, a^-, \vec{a}_{\perp})=((a^0+a^3)/\sqrt{2}, (a^0-a^3)/\sqrt{2}, a^1, a^2)$. The two unit light-cone vectors are denoted as $n^{\mu}=(0,1,\vec{0}_{\perp})$ and $l^{\mu}=(1,0,\vec{0}_{\perp})$. The inner product of four vector $a$ and $b$ then reads $a\cdot b=a^+ b^- + a^- b^+ -\vec{a}_{\perp}\cdot \vec{b}_{\perp}$.

In QCD, the LCDAs are defined by the matrix elements of non-local gauge invariant quark bilinear operators, in which the two fermion fields are separated in the $n$ direction. At the leading twist,
 there are two LCDAs $\phi^{\perp}_{V}$ and $\phi^{\|}_{V}$ corresponding to the transversely (denoted by ``$\perp$'') and longitudinally (denoted by ``$\|$'') polarized states of the vector meson.
We first introduce the non-local operators in coordinate space
\begin{align}
  \mathcal{O}^{\Gamma}_V(\xi^-)=\bar\psi(\xi^-)\Gamma W(\xi^-,0)\psi(0),
\end{align}
where $\Gamma=\gamma^+\gamma^{\alpha}_{\perp}$ for transversely polarized vector meson, and $\Gamma=\gamma^+$ for longitudinally polarized vector meson. $W(\xi^-,0)$ is the Wilson line with the end points $(0,\xi^-,0_{\perp})$ and $(0,0,0_{\perp})$.
In LCDAs the Wilson line is light-like
\begin{align}
  W(\xi^-, 0)=P\exp\bigg[-ig_s\int^{\xi^-}_0 n\cdot A(\lambda n)d\lambda\bigg],
\end{align}
where $P$ denotes that the exponential is path ordered.
 We also need the Fourier transformation of these operators, which are denoted by $O^{\Gamma}_V(x)$
\begin{align}
O^{\Gamma}_V(x)&=\int \frac{d\xi^-}{2\pi}e^{-i x\xi^- P^+}\mathcal{O}^{\Gamma}_V(\xi^-),
\end{align}
$x\equiv k^+/P^+$ is the longitudinal momentum fraction with $k^+$ be the momentum of quark. Then, the LCDAs of the transversely and longitudinally polarized vector meson are defined by the matrix elements of $O^{\Gamma}_V(x)$, in which $O^{\Gamma}_V(x)$ is sandwiched between the meson and vacuum states
\begin{subequations}\label{eq:def:lc}
\begin{align}
  f^{\perp}_V \epsilon^{*\alpha}_{\perp} \phi^{\perp}_{V}(x,\mu)&=\langle V,P,\epsilon^*|O^{\perp}_{V}(x)|0\rangle,\\
   f_V \frac{m_V}{P^+} \epsilon^{*+} \phi^{\|}_{V}(x,\mu)&=\langle V,P,\epsilon^*|O^{\|}_V(x)|0\rangle,
\end{align}
\end{subequations}
where $f^{\perp}_V$ and $f_V$ are the decay constants of the vector meson $V$, $P$ and $\epsilon^*$ are the momentum and polarization vector of meson $V$, respectively.
The decay constants are defined by the local operators
\begin{subequations}\label{eq:local:lc}
\begin{align}
  f^{\perp}_V \epsilon^{*\alpha}_{\perp}&=\langle V,P,\epsilon^*|\mathcal{O}^{\perp}_{V}(0)|0\rangle=\int dx \langle V,P,\epsilon^*|{O}^{\perp}_{V}(x)|0\rangle ,\\
  f_V \frac{m_V}{P^+}\epsilon^{*+} &=\langle V,P,\epsilon^*|\mathcal{O}^{\|}_V(0)|0\rangle=\int dx \langle V,P,\epsilon^*|{O}^{\|}_V(x)|0\rangle,
\end{align}
\end{subequations}
then, the LCDAs can be expressed as the ratio of the non-local and local matrix elements
\begin{align}\label{eq:lc:norm}
  \phi^{\Gamma}_{V}(x,\mu)=\frac{\langle V, P, \epsilon^*|O^{\Gamma}_V(x)|0\rangle}{\langle V,P,\epsilon^*|\mathcal{O}^{\Gamma}_{V}(0)|0\rangle}.
\end{align}

In LaMET, one can define the quasi-DAs similarly.
It is more convenient to discuss quasi observables in the original Cartesian coordinate system. The unit vector of the $z$ direction is denoted by $n^{\mu}_z=(0,0,0,1)$. The inner product of $n_z$ and an arbitrary vector $a$ gives $n_z\cdot a=a_z=-a^z$. To define the quasi-DAs, we introduce two non-local bilinear operators, in which the fermion fields are  separated on the $z$ direction
\begin{align}
  \t{\mathcal{O}}^{\Gamma}(z)=\bar\psi(z)\Gamma W(z,0)\psi(0),
\end{align}
and their Fourier transformation
\begin{align}
\t{O}^{\Gamma}_V(x)&=\int \frac{dz}{2\pi}e^{-i x z P_z}\t{\mathcal{O}}^{\Gamma}(z),
\end{align}
where $\Gamma=\gamma_z\gamma^{\alpha}_{\perp}$ and $\gamma_z$ for the transverse and longitudinal components, respectively. Here the Wilson line $W$ is along the $z$ direction, with  $z n^{\mu}_z=(0,0,0,z)$ and the origin of coordinates $(0,0,0,0)$ as its end points.
Then, the quasi-DAs of the transverse and longitudinal components of a vector meson are defined by the matrix elements of the operators as
\begin{subequations}\label{eq:def:quasi}
\begin{align}
  {f}^{\perp}_V \epsilon^{*\alpha}_{\perp} \t{\phi}^{\perp}_{V}(x, P_z)&=\langle V,P,\epsilon^*|\t{O}^{\perp}_{V}(x)|0\rangle,\\
  {f}_V \epsilon^{*}_z \frac{m_V}{P_z} \t{\phi}^{\|}_{V}(x,P_z)&=\langle V,P,\epsilon^*|\t{O}^{\|}_V(x)|0\rangle,
\end{align}
\end{subequations}
We note that although the light-cone and quasi operators are different, the decay constants in Eqs.~\eqref{eq:def:lc} and \eqref{eq:def:quasi} should be the same. The reason is that either
the quasi or light-cone
operator is the $\mu=+$ or $\mu=z$ component of the operator $\bar\psi \gamma^{\mu}\gamma^{\alpha}_{\perp}\psi$ or $\bar\psi\gamma^{\mu}\psi$. The Lorentz index is only carried by the polarization
vector $\epsilon^*$, while the decay constants are scalar quantities, therefore they are independent of the Lorentz indices.
The decay constants are related to the matrix elements of local operators by
\begin{subequations}\label{eq:local:quasi}
\begin{align}
  {f}^{\perp}_V \epsilon^{*\alpha}_{\perp}&=\langle V,P,\epsilon^*|\t{\mathcal{O}}^{\perp}_{V}(0)|0\rangle=\int dx \langle V,P,\epsilon^*|\t{O}^{\perp}_{V}(x)|0\rangle,\\
  {f}_V \epsilon^{*}_z \frac{m_V}{P_z} &=\langle V,P,\epsilon^*|\t{\mathcal{O}}^{\|}_V(0)|0\rangle=\int dx \langle V,P,\epsilon^*|\t{O}^{\|}_V(x)|0\rangle.
\end{align}
\end{subequations}
The quasi-DA can be expressed as the ratio of the non-local and the local matrix elements as
\begin{align}
  \widetilde{\phi}^{\Gamma}_{V}(x,\mu)=\frac{\langle V, P, \epsilon^*|\t{O}^{\Gamma}_V(x)|0\rangle}{\langle V,P,\epsilon^*|\widetilde{\mathcal{O}}^{\Gamma}_{V}(0)|0\rangle}.
\end{align}
One can immediately find that the LCDAs and quasi-DAs are normalized to 1, i.e.,
\begin{align}\label{eq:norm}
  \int dx \phi^{\Gamma}_V(x,\mu)=1,~~  \int dx \widetilde{\phi}^{\Gamma}_V(x,P_z)=1
\end{align}
from the definitions.

\section{One loop results}\label{sec:oneloop}

To examine the factorization and determine the matching coefficients at one loop level,
we first replace the meson state $\langle V, P,\epsilon^*|$ with its lowest Fock state $\langle Q(x_0 P)\bar Q((1-x_0)P)|$.
 $P$ is the total momentum of the quark and anti-quark, $x_0 P$ and $(1-x_0)P$ are the momenta of the $Q$ and $\bar Q$, respectively, with $0<x_0<1$.
 Then the matrix elements with the Fock state as their final state can be calculated in perturbation theory. Direct calculation at tree level leads to
\begin{align}\label{eq:treelevel}
  \phi^{\Gamma(0)}_V(x)=\t{\phi}^{\Gamma(0)}_V(x)=\delta(x-x_0).
\end{align}

\begin{figure}[hbt]
\begin{center}
\includegraphics[width=0.8\textwidth]{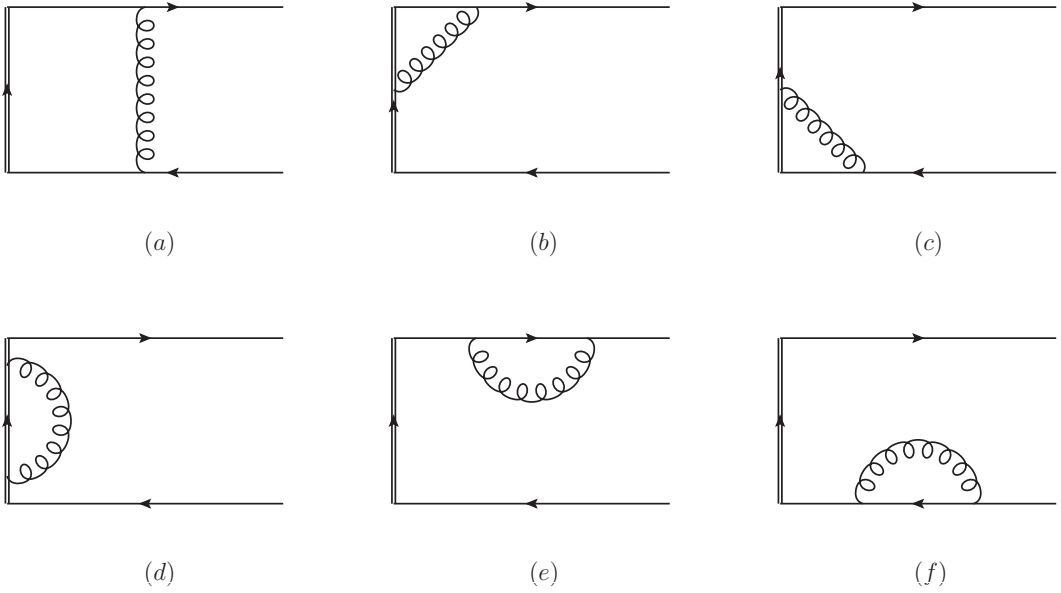}
\end{center}
\caption{Feynman diagrams for LCDAs and quasi-DAs at one loop level. The double line denotes the Wilson line. }
\label{fig:feynman}
\end{figure}

We will perform our calculation under Feynman gauge.
The Feynman diagrams at one loop level are presented by Fig.~\ref{fig:feynman}.
The distribution amplitudes of the Fock state is calculable in perturbation theory, thus can be expanded in series of $\alpha_s$. Up to one loop level, we have
\begin{align}\label{eq:phi:exp}
  \t{\phi}^{\Gamma}_V(x, P_z, \Lambda)=\t{\phi}^{\Gamma(0)}_V(x, P_z)+\t{\phi}^{\Gamma(1)}_V(x, P_z, \Lambda)+\mathcal{O}(\alpha^2_s).
\end{align}
On the other hand,
the matrix element of $\t{O}^{\Gamma}_V(x)$, up to one loop level, can be expressed by
\begin{align}\label{eq:corr:nonlocal}
  \langle V,P,\epsilon^*|\t{O}^{\Gamma}_{V}(x)|0\rangle&=  \langle V,P,\epsilon^*|\t{O}^{\Gamma}_{V}(x)|0\rangle^{(0)}(1+\delta Z^{(1)}_F)+\langle V,P,\epsilon^*|\t{O}^{\Gamma}_{V}(x)|0\rangle^{(1)}+\mathcal{O}(\alpha^2_s).
\end{align}
Here $\langle V,P,\epsilon^*|\t{O}^{\Gamma}_{V}(x)|0\rangle^{(0)}$ is the tree level matrix element, and $\langle V,P,\epsilon^*|\t{O}^{\Gamma}_{V}(x)|0\rangle^{(1)}$ denotes the one loop correction to the matrix element in which the self energy of quark has been excluded. The contributions from quark's self energy are involved in $\delta Z^{(1)}_F$, where $\delta Z^{(1)}_F$ is the one loop correction of  quark's self energy.
Meanwhile, the local matrix element is also corrected at one loop
\begin{align}\label{eq:corr:local}
  \langle V,P,\epsilon^*|\t{\mathcal{O}}^{\Gamma}_{V}(0)|0\rangle&=  \langle V,P,\epsilon^*|\t{\mathcal{O}}^{\Gamma}_{V}(0)|0\rangle^{(0)}(1+\delta Z^{(1)}_F+\delta {Z}^{\Gamma(1)}_V)+\mathcal{O}(\alpha^2_s).
\end{align}
where $\delta {Z}^{\Gamma(1)}_V$ is the one loop vertex correction of the local operator. Since $\mathcal{O}^{\Gamma}_V$ and $\t{\mathcal{O}}^{\Gamma}_V$ are the $\mu=+$ and $\mu=z$ components of
operator $\bar\psi\gamma^{\mu}\gamma^{\alpha}_{\perp}\psi$ or $\bar\psi\gamma^{\mu}\psi$, respectively, $\delta Z^{\Gamma(1)}_V$ should be same for light-cone and quasi local operators.
From Eqs.~\eqref{eq:local:lc} and \eqref{eq:local:quasi}, one can get that
\begin{align}
  \delta {Z}^{\Gamma(1)}_V=\int dx \frac{\langle V,P,\epsilon^*|\t{O}^{\Gamma}_{V}(x)|0\rangle^{(1)}}{\langle V,P,\epsilon^*|\t{\mathcal{O}}^{\Gamma}_{V}(0)|0\rangle^{(0)}}.
\end{align}
From Eqs.~\eqref{eq:corr:nonlocal} and \eqref{eq:corr:local}, we immediately have
\begin{align}\label{eq:da}
  \t{\phi}^{\Gamma}_V(x, \mu)
  &=\frac{\langle V, P, \epsilon^*|\t{O}^{\Gamma}_{V}(x)|0\rangle^{(0)}}{\langle V,P,\epsilon^*|\t{\mathcal{O}}^{\Gamma}_{V}(0)|0\rangle^{(0)}}+\frac{\langle V,P,\epsilon^*|\t{O}^{\Gamma}_{V}(x)|0\rangle^{(1)}}{\langle V,P,\epsilon^*|\t{\mathcal{O}}^{\Gamma}_{V}(0)|0\rangle^{(0)}}-\delta {Z}^{\Gamma(1)}_V \frac{\langle V,P,\epsilon^*|\t{O}^{\Gamma}_{V}(x)|0\rangle^{(0)}}{\langle V,P,\epsilon^*|\t{\mathcal{O}}^{\Gamma}_{V}(0)|0\rangle^{(0)}}+\mathcal{O}(\alpha^2_s)\nonumber\\
    &=\frac{\langle V, P, \epsilon^*|\t{O}^{\Gamma}_{V}(x)|0\rangle^{(0)}}{\langle V,P,\epsilon^*|\t{\mathcal{O}}^{\Gamma}_{V}(0)|0\rangle^{(0)}}+\frac{\langle V,P,\epsilon^*|\t{O}^{\Gamma}_{V}(x)|0\rangle^{(1)}}{\langle V,P,\epsilon^*|\t{\mathcal{O}}^{\Gamma}_{V}(0)|0\rangle^{(0)}}-\frac{\langle V,P,\epsilon^*|\t{O}^{\Gamma}_{V}(x)|0\rangle^{(0)}}{\langle V,P,\epsilon^*|\t{\mathcal{O}}^{\Gamma}_{V}(0)|0\rangle^{(0)}}\int dy \frac{\langle V,P,\epsilon^*|\t{O}^{\Gamma}_{V}(y)|0\rangle^{(1)}}{\langle V,P,\epsilon^*|\t{\mathcal{O}}^{\Gamma}_{V}(0)|0\rangle^{(0)}}\nonumber\\
    &~~~~+\mathcal{O}(\alpha^2_s).
\end{align}
By comparing Eq.~\eqref{eq:da} and Eq.~\eqref{eq:phi:exp}, one can identify that
\begin{align}
  &\t{\phi}^{\Gamma(0)}_V(x)=\frac{\langle V, P, \epsilon^*|\t{O}^{\Gamma}_{V}(x)|0\rangle^{(0)}}{\langle V,P,\epsilon^*|\t{\mathcal{O}}^{\Gamma}_{V}(0)|0\rangle^{(0)}}=\delta(x-x_0),\\
  &\t{\phi}^{\Gamma(1)}_V(x)=\frac{\langle V, P, \epsilon^*|\t{O}^{\Gamma}_{V}(x)|0\rangle^{(1)}}{\langle V,P,\epsilon^*|\t{\mathcal{O}}^{\Gamma}_{V}(0)|0\rangle^{(0)}}-\delta(x-x_0)\int dy \frac{\langle V,P,\epsilon^*|\t{O}^{\Gamma}_{V}(y)|0\rangle^{(1)}}{\langle V,P,\epsilon^*|\t{\mathcal{O}}^{\Gamma}_{V}(0)|0\rangle^{(0)}}.\label{eq:phi1}
\end{align}
From Eq.~\eqref{eq:da}, one can indicates that since quark's self energy cancels between the non-local and local matrix elements, it will not contribute to the distribution amplitudes. Therefore we have no need to consider Fig.~\ref{fig:feynman} (e) and (f). The  general discussions of one loop correction above are also applicable to the LCDAs.

In the following calculations, we will introduce a small gluon mass $m_g$ to regularize the collinear divergence. For the UV divergence,
we will employ two schemes: one is adding an UV cut-off $\Lambda$ on the transverse momentum, another one is the dimensional regularization (DR).
In this section we devote to the cut-off scheme, the DR results will be arranged in Appendix~\ref{sec:DR}.

To express the one loop results of quasi and light cone distribution amplitudes, we introduce the generalized plus distribution ``$+$'', which are defined by
\begin{align}\label{eq:plus}
  \int^{\infty}_{-\infty} d x [f(x)]_{+} T(x)&=\int^{\infty}_{-\infty} dx f(x)(T(x)-T(x_0)),
\end{align}
where $T(x)$ is an arbitrary smooth test function. The generalized plus function regularizes the pole of divergent integral at $x=x_0$.

\subsection{Transverse distribution amplitudes}

We now list the results of distribution amplitudes for the transversely polarized vector meson diagram by diagram.

In Fig.~\ref{fig:feynman}(a), the internal gluon is not connected to the Wilson line. For this diagram, we have
\begin{align}
  \phi^{\perp(1)}_V(x, \Lambda)\bigg\vert_{\mathrm{Fig}.~\ref{fig:feynman}(a)}=\t{\phi}^{\perp(1)}_V(x, P_z, \Lambda)\bigg\vert_{\mathrm{Fig}.~\ref{fig:feynman}(a)}=0
\end{align}
for both LCDA and quasi-DA.

In Fig.~\ref{fig:feynman}(b)(c), one end of the internal gluon is attached to the Wilson line, thus there is an eikonal propagator, which is proportional to $1/(x-x_0)$. The contributions from Fig.~\ref{fig:feynman}(b) read
\begin{align}
  \phi^{\perp(1)}_V(x, \Lambda)\bigg\vert_{\mathrm{Fig}.~\ref{fig:feynman}(b)}=\frac{\alpha_s C_F}{2\pi}\left\{\begin{aligned}
  &\bigg[\frac{x}{x_0(x-x_0)}\ln\frac{m^2_g x}{\Lambda^2 x_0}\bigg]_{+}, &0<x<x_0\\
  &0\,, &\mathrm{others}
  \end{aligned}\right.
\end{align}
for LCDA, and
\begin{align}
  \t{\phi}^{\perp(1)}_V(x, P_z, \Lambda)\bigg\vert_{\mathrm{Fig}.~\ref{fig:feynman}(b)}=\frac{\alpha_s C_F}{2\pi}  \left\{\begin{aligned}
  &\bigg[\frac{x}{x_0(x-x_0)}\ln\frac{x}{x-x_0}-\frac{1}{2(x-x_0)}\bigg]_{+},~~~~&x<0\\
  &\bigg[\frac{x}{x_0(x-x_0)}\ln\frac{m^2_g}{4 P^2_z x_0(x_0-x)}+\frac{2x-x_0}{2x_0(x-x_0)}\bigg]_{+},~~~~&0<x<x_0\\
  &\bigg[\frac{x}{x_0(x-x_0)}\ln\frac{x-x_0}{x}+\frac{1}{2(x-x_0)}\bigg]_+,~~~~&x>x_0
  \end{aligned}
    \right.
\end{align}
for quasi-DA.
For Fig.~\ref{fig:feynman}(c), we have
\begin{align}
  \phi^{\perp(1)}_V(x, \Lambda)\bigg\vert_{\mathrm{Fig}.~\ref{fig:feynman}(c)}=\frac{\alpha_s C_F}{2\pi}\left\{\begin{aligned}
  &-\bigg[\frac{x-1}{(x_0-1)(x-x_0)}\ln\frac{m^2_g (x-1)}{\Lambda^2 (x_0-1)}\bigg]_+, &x_0<x<1\\
  &0\,, &\mathrm{others} \end{aligned}\right.
\end{align}
for LCDA, and
\begin{align}
  \t{\phi}^{\perp(1)}_V(x, P_z, \Lambda)\bigg\vert_{\mathrm{Fig}.~\ref{fig:feynman}(c)}=\frac{\alpha_s C_F}{2\pi}\left\{\begin{aligned}
    &\bigg[\frac{x-1}{(x_0-1)(x-x_0)}\ln\frac{x-1}{x-x_0}-\frac{1}{2(x-x_0)}\bigg]_{+},~~~~&x<x_0\\
    &\bigg[-\frac{x-1}{(x_0-1)(x-x_0)}\ln\frac{m^2_g}{4P^2_z(1-x_0)(x-x_0)}\\
    &-\frac{2x-x_0-1}{2(x_0-1)(x-x_0)}\bigg]_+,~~~~&x_0<x<1\\
    &\bigg[\frac{x-1}{(x_0-1)(x-x_0)}\ln\frac{x-x_0}{x-1}+\frac{1}{2(x-x_0)}\bigg]_{+},~~~~&x>1
  \end{aligned}\right.
\end{align}
for quasi-DA.

Fig.~\ref{fig:feynman}(d) is the one loop correction to Wilson line's self energy, which is proportional to $n^2$, $n$ is the direction vector of Wilson line. This contribution vanishes for LCDA since $n^2=0$, but do not vanish for quasi-DA.
 Then the results read
\begin{align}
  \phi^{\perp(1)}_V(x, \Lambda)\bigg\vert_{\mathrm{Fig}.~\ref{fig:feynman}(d)}=0\,,
\end{align}
and
\begin{align}\label{eq:Wilsonselfenergy}
  \t{\phi}^{\perp(1)}_V(x, P_z, \Lambda)\bigg\vert_{\mathrm{Fig}.~\ref{fig:feynman}(d)}=\frac{\alpha_s C_F}{2\pi}\left\{\begin{aligned}
    &\bigg[\frac{1}{x-x_0}+\frac{\Lambda}{(x-x_0)^2P_z}\bigg]_{+},~~~~&x<x_0\\
    &\bigg[-\frac{1}{x-x_0}+\frac{\Lambda}{(x-x_0)^2P_z}\bigg]_+.~~~~&x>x_0
  \end{aligned}\right.
\end{align}
Note that for quasi-DA, this diagram contributes a linear divergence.
 Perturbative calculation on quasi-PDFs also shows the existence of the power-like UV divergence~\cite{Xiong:2013bka,Ji:2015qla,Wang:2017qyg}. The power divergences have to be subtracted properly. A renormalization scheme has been proposed to subtract the linear divergence based on the auxiliary field formalism~\cite{Chen:2016fxx,Ji:2018hvs,Ishikawa:2016znu,Wang:2017eel};
 another approach is to replace the straight Wilson line with the non-dipolar Wilson lines~\cite{Li:2016amo}.

 It has been known that the power divergence in Wilson line's self energy can be canceled by introducing a ``mass counter term'' of the Wilson line~\cite{Polyakov:1980ca}. Since the source of the linear divergence is the Wilson line's self energy, the improved quasi-PDFs and DAs are proposed by adding such a mass counter term, to subtract the linear divergence~\cite{Chen:2016fxx,Zhang:2017bzy}. In the same spirit, one can also define the improved quasi-DAs of vector meson. To do this, we replace the
operator in Eq.~\eqref{eq:def:quasi} by the ``improved'' operator
\begin{align}\label{eq:imp}
\t{O}^{\Gamma~\mathrm{imp.}}_V(x)&=\int \frac{dz}{2\pi}e^{-i x z P_z-\delta m|z|}\t{\mathcal{O}}^{\Gamma}(z),
\end{align}
where $\delta m$ is the mass counter term of the Wilson line. It has been shown that $\delta m$ can be extracted by using the static quark potential non-perturbatively~\cite{Musch:2010ka}. Perturbative calculation shows that the contribution from $\delta m$ cancels the linearly divergent term in Eq.~\eqref{eq:Wilsonselfenergy}. Therefore, one can get the result for improved quasi-DAs just by subtracting the linearly divergent term.

In the above results, the LCDAs are only non-zero in the physical regions $0<x<x_0$ and $x_0<x<1$, while the quasi-DAs have non-zero support in all of the four regions $x<0$, $0<x<x_0$, $x_0<x<1$ and $x>1$. However, the collinear divergence only exists in the physical regions $0<x<x_0$ and $x_0<x<1$.
One can also notice that the LCDAs and quasi-DAs are symmetric under variable substitution $x\leftrightarrow 1-x,~x_0\leftrightarrow 1-x_0$.

\subsection{Longitudinal distribution amplitudes}

The one loop results of distribution amplitudes for longitudinally polarized vector meson are listed below diagram by diagram.

For Fig.~\ref{fig:feynman}(a), we have
\begin{align}
  \phi^{\|(1)}_V(x, \Lambda)\bigg\vert_{\mathrm{Fig}.~\ref{fig:feynman}(a)}=\frac{\alpha_s C_F}{2\pi}\left\{\begin{aligned}
    &\bigg[-\frac{x}{x_0}\ln\frac{m^2_g x}{\Lambda^2 x_0}\bigg]_+,~~~~&0<x<x_0\\
    &\bigg[-\frac{x-1}{x_0-1}\ln\frac{m^2(x-1)}{\Lambda^2(x_0-1)}\bigg]_+,~~~~&x_0<x<1\\
    &0,~~~~&\mathrm{others}
  \end{aligned}\right.
\end{align}
for LCDA,
and
\begin{align}
  \t{\phi}^{\|(1)}_V(x, P_z, \Lambda)\bigg\vert_{\mathrm{Fig}.~\ref{fig:feynman}(a)}=\frac{\alpha_s C_F}{2\pi}\left\{\begin{aligned}
    &\bigg[\frac{x-1}{x_0-1}\ln\frac{x-1}{x-x_0}-\frac{x}{x_0}\ln\frac{x}{x-x_0}\bigg]_{+},~~~~&x<0\\
    &\bigg[-\frac{x}{x_0}\ln\frac{m^2_g}{4 P^2_z x_0(x_0-x)}-\frac{x-1}{x_0-1}\ln\frac{x-x_0}{x-1}\bigg]_+,~~~~&0<x<x_0\\
    &\bigg[-\frac{x-1}{x_0-1}\ln\frac{m^2_g}{4P^2_z(x_0-1)(x_0-x)}-\frac{x}{x_0}\ln\frac{x-x_0}{x}\bigg]_+,~~~~&x_0<x<1\\
    &\bigg[\frac{x}{x_0}\ln\frac{x}{x-x_0}-\frac{x-1}{x_0-1}\ln\frac{x-1}{x-x_0}\bigg]_{+},&~~~~x>1
  \end{aligned}\right.
\end{align}
for quasi-DA. Note that according to Eq.~\eqref{eq:phi1}, we have subtract the vertex correction of the local operator which can be expressed as an integral of the non-local matrix element.
Therefore the contributions above have been reformed to the generalized plus distribution.

For Fig.~\ref{fig:feynman}(b), the results are
\begin{align}
  \phi^{\|(1)}_V(x, \Lambda)\bigg\vert_{\mathrm{Fig}.~\ref{fig:feynman}(b)}=\frac{\alpha_s C_F}{2\pi}\left\{\begin{aligned}
  &\bigg[\frac{x}{x_0(x-x_0)}\ln\frac{m^2_g x}{\Lambda^2 x_0}\bigg]_{+}, &0<x<x_0\\
  &0, &\mathrm{others}
  \end{aligned}
  \right.
\end{align}
for LCDA, and
\begin{align}
  \t{\phi}^{\|(1)}_V(x, P_z, \Lambda)\bigg\vert_{\mathrm{Fig}.~\ref{fig:feynman}(b)}=\frac{\alpha_s C_F}{2\pi}  \left\{\begin{aligned}
  &\bigg[\frac{x}{x_0(x-x_0)}\ln\frac{x}{x-x_0}-\frac{1}{2(x-x_0)}\bigg]_{+},~~~~&x<0\\
  &\bigg[\frac{x}{x_0(x-x_0)}\ln\frac{m^2_g}{4 P^2_z x_0(x_0-x)}+\frac{2x-x_0}{2x_0(x-x_0)}\bigg]_{+},~~~~&0<x<x_0\\
  &\bigg[\frac{x}{x_0(x-x_0)}\ln\frac{x-x_0}{x}+\frac{1}{2(x-x_0)}\bigg]_+,~~~~&x>x_0
  \end{aligned}
    \right.
\end{align}
for quasi-DA. Similarly, for  Fig.~\ref{fig:feynman}(c), we have
\begin{align}
  \phi^{\|(1)}_V(x, \Lambda)\bigg\vert_{\mathrm{Fig}.~\ref{fig:feynman}(c)}=\frac{\alpha_s C_F}{2\pi}\left\{\begin{aligned}
  &-\bigg[\frac{x-1}{(x_0-1)(x-x_0)}\ln\frac{m^2_g (x-1)}{\Lambda^2 (x_0-1)}\bigg]_+, &x_0<x<1\\
  &0, &\mathrm{others}\end{aligned}\right.
\end{align}
for LCDA, and
\begin{align}
  \t{\phi}^{\|(1)}_V(x, P_z, \Lambda)\bigg\vert_{\mathrm{Fig}.~\ref{fig:feynman}(c)}=\frac{\alpha_s C_F}{2\pi}\left\{\begin{aligned}
    &\bigg[\frac{x-1}{(x_0-1)(x-x_0)}\ln\frac{x-1}{x-x_0}-\frac{1}{2(x-x_0)}\bigg]_{+},~~~~&x<x_0\\
    &\bigg[-\frac{x-1}{(x_0-1)(x-x_0)}\ln\frac{m^2_g}{4P^2_z(1-x_0)(x-x_0)}\\
    &-\frac{2x-x_0-1}{2(x_0-1)(x-x_0)}\bigg]_+,~~~~&x_0<x<1\\
    &\bigg[\frac{x-1}{(x_0-1)(x-x_0)}\ln\frac{x-x_0}{x-1}+\frac{1}{2(x-x_0)}\bigg]_{+},~~~~&x>1
  \end{aligned}\right.
\end{align}
for quasi-DA.

Fig.~\ref{fig:feynman}(d) receives contribution from Wilson line's self energy, which is proportional to $n^2$, $n$ is the direction vector of Wilson line. This contribution vanishes for LCDAs since $n^2=0$, but does not vanish for quasi-DAs. The results reads
\begin{align}
  \phi^{\|(1)}_V(x, \Lambda)\bigg\vert_{\mathrm{Fig}.~\ref{fig:feynman}(d)}=0,
\end{align}
for LCDA, and
\begin{align}\label{eq:Wilsonselfenergy:long}
  \t{\phi}^{\|(1)}_V(x, P_z, \Lambda)\bigg\vert_{\mathrm{Fig}.~\ref{fig:feynman}(d)}=\frac{\alpha_s C_F}{2\pi}\left\{\begin{aligned}
    &\bigg[\frac{1}{x-x_0}+\frac{\Lambda}{(x-x_0)^2 P_z}\bigg]_{+},~~~~&x<x_0\\
    &\bigg[-\frac{1}{x-x_0}+\frac{\Lambda}{(x-x_0)^2 P_z}\bigg]_+,~~~~&x>x_0\\
  \end{aligned}\right.
\end{align}
is the result of quasi-DA.
Similar to the transverse quasi-DA, this diagram also contributes a linear divergence to the longitudinal quasi-DA. As we have discussed in the last subsection, the linear divergence can be cured by introducing a mass counter term of Wilson line. The improved quasi-DAs have already been defined by Eq.~\eqref{eq:imp}. The one loop results under the improved definition can be got by subtracting the linearly divergent term in Eq.~\eqref{eq:Wilsonselfenergy:long}.

At last, since that all of the results above are represented by the generalized plus distribution, thus they are zero under the integration, which is the normalization condition given by Eq.~\eqref{eq:norm}.

\section{The matching equation}\label{sec:match}

In this section, we present the matching equation connecting the LCDAs and quasi-DAs.

In LaMET, if the factorization holds, the quasi-DA $\t{\phi}^{\Gamma}_V$ can be factorized as
\begin{align}\label{eq:fac}
  \t{\phi}^{\Gamma}_V(x, P_z, \Lambda)=\int^{1}_0 dy Z_{\Gamma}(x, y, P_z, \Lambda )\phi^{\Gamma}_V(y,\Lambda)+\mathcal{O}\bigg(\frac{\Lambda^2_{\mathrm{QCD}}}{P^2_z},\frac{m^2_V}{P^2_z}\bigg),
\end{align}
where $y$ is constrained by $0<y<1$.
Here $Z_{\Gamma}$ is the perturbatively calculable function, hence can be expanded in the series of $\alpha_s$ as
\begin{align}
  Z_{\Gamma}(x,y,P_z,\Lambda)&=\sum^{\infty}_{n=0}\bigg(\frac{\alpha_s}{2\pi}\bigg)^n Z^{(n)}_{\Gamma}(x,y,P_z,\Lambda)\nonumber\\
  &=\delta(x-y)+\frac{\alpha_s}{2\pi}Z^{(1)}_{\Gamma}(x,y,P_z,\Lambda)+\mathcal{O}(\alpha^2_s).
\end{align}
By recalling the tree level result in Eq.~\eqref{eq:treelevel}, one can find that the one loop correction to the matching coefficient can be attributed to the difference between LCDA and quasi-DA at one loop level,
\begin{align}\label{eq:match}
  \frac{\alpha_s}{2\pi}Z^{(1)}_{\Gamma}(x,x_0,P_z,\Lambda)=\t{\phi}^{\Gamma(1)}_V(x, P_z, \Lambda)-\phi^{\Gamma(1)}_V(x, \Lambda).
\end{align}
By using Eq.~\eqref{eq:match}, together with the one loop results calculated in Sec.~\ref{sec:oneloop}, one can determine the one loop corrections to the matching coefficients.
For $Z^{(1)}_{\perp}$, we have
\begin{align}\label{eq:match:trans}
  Z^{(1)}_{\perp}(x,y,P_z,\Lambda)
  =C_F\left\{\begin{aligned}
    &\bigg[\frac{x}{y(x-y)}\ln\frac{x}{x-y}+\frac{x-1}{(y-1)(x-y)}\ln\frac{x-1}{x-y}\\
    &+\frac{\Lambda}{(x-y)^2 P_z}\bigg]_{+}, &x<0<y\\
    &\bigg[\frac{x}{y(x-y)}\ln\frac{\Lambda^2}{4P^2_z x(y-x)}+\frac{x-1}{(y-1)(x-y)}\ln\frac{x-1}{x-y}+\frac{x}{y(x-y)}\\
    &+\frac{\Lambda}{(x-y)^2 P_z}\bigg]_+, &0<x<y\\
    &\bigg[-\frac{x-1}{(y-1)(x-y)}\ln\frac{\Lambda^2}{4P^2_z(1-x)(x-y)}+\frac{x}{y(x-y)}\ln\frac{x-y}{x}\\
    &+\frac{1-x}{(y-1)(x-y)}+\frac{\Lambda}{(x-y)^2 P_z}\bigg]_+, &y<x<1\\
    &\bigg[\frac{x}{y(x-y)}\ln\frac{x-y}{x}+\frac{x-1}{(y-1)(x-y)}\ln\frac{x-y}{x-1}\\
    &+\frac{\Lambda}{(x-y)^2 P_z} \bigg]_{+}, &y<1<x
  \end{aligned}\right.
\end{align}
and
for $Z^{(1)}_{\|}$, the result reads
\begin{align}\label{eq:match:long}
  Z^{(1)}_{\|}(x,y,P_z,\Lambda)
  =C_F\left\{\begin{aligned}
    &\bigg[\frac{x-1}{y-1}\bigg(1+\frac{1}{x-y}\bigg)\ln\frac{x-1}{x-y}-\frac{x}{y}\bigg(1-\frac{1}{x-y}\bigg)\ln\frac{x}{x-y}\\
    &+\frac{\Lambda}{(x-y)^2 P_z}\bigg]_{+}, &x<0<y\\
    &\bigg[\frac{x}{y}\bigg(1-\frac{1}{x-y}\bigg)\ln\frac{4P^2_z(y-x)x}{\Lambda^2}-\frac{x-1}{y-1}\bigg(1+\frac{1}{x-y}\bigg)\ln\frac{x-y}{x-1}\\
    &-\frac{x}{y(x-y)}+\frac{\Lambda}{(x-y)^2 P_z}\bigg]_+, &0<x<y\\
    &\bigg[\frac{x-1}{y-1}\bigg(1+\frac{1}{x-y}\bigg)\ln\frac{4P^2_z(1-x)(x-y)}{\Lambda^2}-\frac{x}{y}\bigg(1-\frac{1}{x-y}\bigg)\ln\frac{x-y}{x}\\
    &+\frac{1-x}{(y-1)(x-y)}+\frac{\Lambda}{(x-y)^2 P_z}\bigg]_+, &y<x<1\\
    &\bigg[\frac{x}{y}\bigg(1-\frac{1}{x-y}\bigg)\ln\frac{x}{x-y}-\frac{x-1}{y-1}\bigg(1+\frac{1}{x-y}\bigg)\ln\frac{x-1}{x-y}\\
    &+\frac{\Lambda}{(x-y)^2 P_z}\bigg]_{+}. &y<1<x
  \end{aligned}\right.
\end{align}
In other regions, $Z^{(1)}_{\perp}$ and $Z^{(1)}_{\|}$ are zero. One can notice that $Z_{\Gamma}(x,y,P_z,\Lambda)=Z_{\Gamma}(1-x,1-y,P_z,\Lambda)$.
We should note that the plus distribution here is to subtract the singularities located at $x=y$,
which is a little different from the one defined in Eq.~\eqref{eq:plus}.
One can immediately find that the collinear divergence, which is regularized by $m_g$, canceled out between LCDAs and quasi-DAs, thus the matching coefficients are free of IR divergence.
Thus we have proved the LaMET factorization for DAs of vector meson at one loop level.

 There are also UV divergence which are regularized by the cut-off $\Lambda$. As we have discussed in Sec.~\ref{sec:oneloop}, the linear divergence will be subtracted by introducing $\delta m$, the mass counter term of Wilson line. Therefore, the matching coefficients of LCDAs and the improved quasi-DAs are the same to Eqs.~\eqref{eq:match:trans} and \eqref{eq:match:long} except the linearly divergent terms, hence the improved matching coefficients have only the logarithm UV divergence. The relation between improved matching coefficients and Eqs.~\eqref{eq:match:trans}\eqref{eq:match:long} is given by
\begin{align}
  Z^{(1),\mathrm{imp.}}_{\Gamma}(x,y,P_z,\Lambda)=Z^{(1)}_{\Gamma}(x,y,P_z,\Lambda)-C_F\bigg[\frac{\Lambda}{(x-y)^2P_z}\bigg]_+.
\end{align}

In Sec.~\ref{sec:oneloop}, we have taken the $\Lambda\to\infty$ limit,
the $\mathcal{O}(P_z/\Lambda)$ contributions have been neglected. However, at present it is difficult to take too large value of $P_z$ in lattice simulations, in fact, $\Lambda$ and $x P_z$ are of 
the same order in a practical calculation on the lattice. Therefore, it is valuable to consider the finite $\Lambda$ corrections to the matching coefficients. The matching coefficients with a finite cut-off have been
derived for the quark PDF~\cite{Xiong:2013bka,Chen:2016utp} and LCDA of pion~\cite{Zhang:2017bzy}. In Appendix~\ref{app:finitelambda} we will list the one loop matching coefficients of vector meson's distribution
amplitudes with a finite cut-off $\Lambda$.

Since LCDAs do not depend on $P_z$, one can take derivative with $\ln P_z$ on both sides of the factorization formula Eq.~\eqref{eq:fac}, and derive the evolution equation of quasi-DAs with $P_z$
\begin{align}
  \frac{d \t{\phi}^{\Gamma,\mathrm{imp.}}_{V}(x, P_z)}{d\ln P_z}=\frac{\alpha_s C_F}{\pi}\int dy V_{\Gamma}(x,y) ~\t{\phi}^{\Gamma,\mathrm{imp.}}_V(y, P_z),
\end{align}
where $V_{\Gamma}(x,y)=d\ln Z_{\Gamma}(x,y,P_z,\Lambda)/d\ln P_z$ is the evolution kernel, and the superscript ``imp.'' denotes that the quasi-DAs are under the improved definition. With the $Z_{\Gamma}$ calculated in the above, we arrive at
\begin{subequations}
\begin{align}
  V_{\perp}(x,y)&=\bigg[\frac{x}{y(y-x)}\theta(y-x)\theta(x)\bigg]_+ +\bigg[\frac{1-x}{(1-y)(x-y)}\theta(x-y)\theta(1-x)\bigg]_+ ,\\
  V_{\|}(x,y)&=\bigg[\frac{x}{y}\bigg(1-\frac{1}{x-y}\bigg)\theta(y-x)\theta(x)\bigg]_+ +\bigg[\frac{1-x}{1-y}\bigg(1+\frac{1}{x-y}\bigg)\theta(x-y)\theta(1-x)\bigg]_+ ,
\end{align}
\end{subequations}
where $\theta(x)$ is the Heaviside step function.
These functions are the Brodsky-Lepage kernels. It indicates that the evolution of quasi-DAs with $P_z$ shares
 the same behavior with the scale evolution of LCDAs, which are dominated by the Efremov-Radyushkin-Brodsky-Lepage (ERBL)
  equation~\cite{Lepage:1979zb,Lepage:1980fj,Efremov:1979qk,Efremov:1978rn}. This evolution equation can be used to resum the large logarithm
  of $P_z$ which appears in the perturbative calculations. The $P_z$ evolution behavior for quasi-PDFs have already been reported, see, e.g., Refs.~\cite{Ji:2013dva,Wang:2017qyg}.
Since the $P_z$ evolution equation of quasi-DAs is equivalent to the ERBL equation of LCDAs,
one can expect that when $P_z\to\infty$, the quasi-DAs converge to the same asymptotic form with LCDAs.
Therefore, it seems that the asymptotic form is the UV fixed point for both LCDAs and quasi-DAs.

\section{summary}
In the framework of large momentum effective theory, we have performed one loop calculation on the leading twist light-cone distribution amplitudes as well as the quasi distribution amplitudes of the vector meson. The distribution amplitudes of both transversely and longitudinally polarized meson have been discussed.
Based on the perturbative calculation under UV cut-off and DR schemes,
we have examined the LaMET factorization and found that the collinear divergence cancels between light-cone and quasi distribution amplitudes. The matching coefficients have been determined at one loop accuracy.
We also get the meson momentum evolution equation for quasi distribution amplitudes, and find that the evolution kernels are identical with the Brodsky-Lepage kernels of light-cone distribution amplitudes.
The results of the present work will be useful to extract light-cone distribution amplitudes of vector mesons from the future lattice simulations.

For practical simulation on the lattice, the renormalization of quasi-DAs is necessary. In the present work the calculation is performed in a naive cut-off scheme
 and the renormalization is absent. Furthermore, the one loop calculation is not on the discrete but the continuum quasi-PDFs. Therefore, a calculation based on lattice perturbation theory, is necessary to fill the gap. Another approach is to renormalize the quasi-DAs in a nonperturbative renormalization scheme, such as the RI/MOM scheme, which has been employed to renormalize quasi-PDFs on the lattice. These issues will be discussed in the future works.

\section*{Acknowledgments}

We  are grateful to Prof.~Wei Wang for critical reading of the manuscript, and his suggestions which inspired this work and improved the presentation of the manuscript greatly. We are also thankful
to Dr.~Yong Zhao for inspiring discussions.
This work is supported  in part  by  National  Natural
Science Foundation of China under Grant
 No.11575110, 11655002, 11735010, 11521505, 11621131001, by Natural Science Foundation of Shanghai under Grant  No.~15DZ2272100 and No.~15ZR1423100, Shanghai Key Laboratory for Particle Physics and Cosmology, and  by  MOE  Key Laboratory for Particle Physics, Astrophysics and Cosmology.

\appendix

\section{One loop results in dimensional regularization}\label{sec:DR}

In Sec.~\ref{sec:oneloop}, we have introduced a cut-off $\Lambda$ on the transverse momentum as an UV regulator. A commonly used regularization scheme is the dimensional regularization. In this scheme, the space-time dimensions are modified from 4 to $d=4-2\epsilon$. The UV divergence is expressed by the poles of $\epsilon$. To renormalize the UV divergence one can employ the $\overline{\mathrm{MS}}$ scheme, in which only the terms proportional to $1/\epsilon-\gamma_E+\ln 4\pi$ ($\gamma_E=0.577...$ is the Euler--Mascheroni constant) are subtracted. Since the standard light-cone PDFs and LCDAs are always defined under $\overline{\mathrm{MS}}$, we list here the one loop results under DR and $\overline{\mathrm{MS}}$.

\subsection{Transverse distribution amplitudes}

We list here our results of distribution amplitudes for transversely polarized vector meson under dimensional regularization.
For Fig.~\ref{fig:feynman}(a), we have
\begin{align}
  \phi^{\perp(1)}_V(x, \mu)\bigg\vert_{\mathrm{Fig}.~\ref{fig:feynman}(a)}=0,
\end{align}
and
\begin{align}
  \t{\phi}^{\perp(1)}_V(x, P_z)\bigg\vert_{\mathrm{Fig}.~\ref{fig:feynman}(a)}=0.
\end{align}

For Fig.~\ref{fig:feynman}(b), we have
\begin{align}
  \phi^{\perp(1)}_V(x, \mu)\bigg\vert_{\mathrm{Fig}.~\ref{fig:feynman}(b)}=-\frac{\alpha_s C_F}{2\pi}\left\{\begin{aligned}
  &\bigg[\frac{x}{x_0(x-x_0)}\ln\frac{\mu^2 x_0}{m^2_g x}\bigg]_{+}, &0<x<x_0,\\
  &0, &\mathrm{others}
  \end{aligned}\right.
\end{align}
\begin{align}
  \t{\phi}^{\perp(1)}_V(x, P_z)\bigg\vert_{\mathrm{Fig}.~\ref{fig:feynman}(b)}=\frac{\alpha_s C_F}{2\pi}  \left\{\begin{aligned}
  &\bigg[\frac{x}{x_0(x-x_0)}\ln\frac{x}{x-x_0}-\frac{1}{2(x-x_0)}\bigg]_{+},~~~~&x<0\\
  &\bigg[\frac{x}{x_0(x-x_0)}\ln\frac{m^2_g}{4 P^2_z x_0(x_0-x)}+\frac{2x-x_0}{2x_0(x-x_0)}\bigg]_{+},~~~~&0<x<x_0\\
  &\bigg[\frac{x}{x_0(x-x_0)}\ln\frac{x-x_0}{x}+\frac{1}{2(x-x_0)}\bigg]_+.~~~~&x>x_0
  \end{aligned}
    \right.
\end{align}

For Fig.~\ref{fig:feynman}(c), we have
\begin{align}
  \phi^{\perp(1)}_V(x, \mu)\bigg\vert_{\mathrm{Fig}.~\ref{fig:feynman}(c)}=\frac{\alpha_s C_F}{2\pi}\left\{\begin{aligned}
  &\bigg[\frac{x-1}{(x_0-1)(x-x_0)}\ln\frac{\mu^2 (1-x_0)}{m^2_g (1-x)}\bigg]_+, &x_0<x<1\\
  &0, &\mathrm{others}
  \end{aligned}\right.
\end{align}
\begin{align}
  \t{\phi}^{\perp(1)}_V(x,P_z)\bigg\vert_{\mathrm{Fig}.~\ref{fig:feynman}(c)}=\frac{\alpha_s C_F}{2\pi}\left\{\begin{aligned}
    &\bigg[\frac{x-1}{(x_0-1)(x-x_0)}\ln\frac{x-1}{x-x_0}-\frac{1}{2(x-x_0)}\bigg]_{+},~~~~&x<x_0\\
    &\bigg[-\frac{x-1}{(x_0-1)(x-x_0)}\ln\frac{m^2_g}{4P^2_z(1-x_0)(x-x_0)}\\
    &-\frac{2x-x_0-1}{2(x_0-1)(x-x_0)}\bigg]_+,~~~~&x_0<x<1\\
    &\bigg[\frac{x-1}{(x_0-1)(x-x_0)}\ln\frac{x-x_0}{x-1}+\frac{1}{2(x-x_0)}\bigg]_{+}.~~~~&x>1
  \end{aligned}\right.
\end{align}

Fig.~\ref{fig:feynman}(d) is the self energy of wilson line. For a Wilson line along the light-cone direction, the self energy is zero. For a space like Wilson line, the self energy is linearly divergent. However, in DR scheme, one can assign a finite value to the linearly divergent self energy with analytical continuation. Thus we have,
\begin{align}
  \phi^{\perp(1)}_V(x, \mu)\bigg\vert_{\mathrm{Fig}.~\ref{fig:feynman}(d)}=0,
\end{align}
\begin{align}
  \t{\phi}^{\perp(1)}_V(x, \mu)\bigg\vert_{\mathrm{Fig}.~\ref{fig:feynman}(d)}=\frac{\alpha_s C_F}{2\pi}\left\{\begin{aligned}
    &\bigg[\frac{1}{x-x_0}\bigg]_{+},~~~~&x<x_0\\
    &\bigg[-\frac{1}{x-x_0}\bigg]_+.~~~~&x>x_0
  \end{aligned}\right.
\end{align}
In the results of LCDAs, we have performed the $\overline{\mathrm{MS}}$ subtraction. For the quasi-DAs, the results of all the one loop diagrams are finite. However, one
can notice that when $x\to\pm\infty$, the quasi-DA behaves as $\propto 1/x$, which is logarithmically divergent. One can take the convolution of the quasi-DA and an arbitrary test funtion $T(x)$, e.g., $T(x)=1$. The integral
is zero since the quasi-DAs is of type $[f(x)]_+$, but it is due to the cancelation of two logarithmically divergent integrals. Thus a renormalization is needed to make the integrals converge.
One calculation on the quasi-PDF based on RI/MOM scheme has been performed in Ref.~\cite{Stewart:2017tvs}. The renormalization on quasi-DAs will be discussed in a forthcoming work.

\subsection{Longitudinal distribution amplitudes}
Now we list our results for the distribution amplitudes of longitudinally polarized vector meson under dimensional regularization.

For Fig.~\ref{fig:feynman}(a), we have
\begin{align}
  \phi^{\|(1)}_V(x, \mu)\bigg\vert_{\mathrm{Fig}.~\ref{fig:feynman}(a)}=\frac{\alpha_s C_F}{2\pi}\left\{\begin{aligned}
    &\bigg[\frac{x}{x_0}\left(\ln\frac{\mu^2 x_0}{m^2_g x}-1\right)\bigg]_+,~~~~&0<x<x_0\\
    &\bigg[\frac{x-1}{x_0-1}\left(\ln\frac{\mu^2(x_0-1)}{m^2_g(x-1)}-1\right)\bigg]_+,~~~~&x_0<x<1\\
    &0,~~~~&\mathrm{others}
  \end{aligned}\right.
\end{align}
and
\begin{align}
  \t{\phi}^{\|(1)}_V(x, P_z)\bigg\vert_{\mathrm{Fig}.~\ref{fig:feynman}(a)}=\frac{\alpha_s C_F}{2\pi}\left\{\begin{aligned}
    &\bigg[\frac{x-1}{x_0-1}\ln\frac{x-1}{x-x_0}-\frac{x}{x_0}\ln\frac{x}{x-x_0}\bigg]_{+},~~~~&x<0\\
    &\bigg[-\frac{x}{x_0}\ln\frac{m^2_g}{4 P^2_z x_0(x_0-x)}-\frac{x-1}{x_0-1}\ln\frac{x-x_0}{x-1}\bigg]_+,~~~~&0<x<x_0\\
    &\bigg[-\frac{x-1}{x_0-1}\ln\frac{m^2_g}{4P^2_z(x_0-1)(x_0-x)}-\frac{x}{x_0}\ln\frac{x-x_0}{x}\bigg]_+,~~~~&x_0<x<1\\
    &\bigg[\frac{x}{x_0}\ln\frac{x}{x-x_0}-\frac{x-1}{x_0-1}\ln\frac{x-1}{x-x_0}\bigg]_{+}.&~~~~x>1
  \end{aligned}\right.
\end{align}
According to Eq.~\eqref{eq:da}, there is a contribution from the vertex correction of the local operator, which can be expressed as a integral of $\phi^{\Gamma}_V(x)$ or $\t{\phi}^{\Gamma}_V(x)$.
Note that we have added the contribution from $\delta Z^{\|(1)}_V\delta(x-x_0)$ here, so the contributions above have been reformed to the generalized plus distribution.

For Fig.~\ref{fig:feynman}(b), we have
\begin{align}
  \phi^{\|(1)}_V(x, \mu)\bigg\vert_{\mathrm{Fig}.~\ref{fig:feynman}(b)}=-\frac{\alpha_s C_F}{2\pi}\left\{\begin{aligned}
  &\bigg[\frac{x}{x_0(x-x_0)}\ln\frac{\mu^2 x_0}{m^2_g x}\bigg]_+, &0<x_0<x\\
  &0, &\mathrm{others}\end{aligned}\right.
\end{align}
\begin{align}
  \t{\phi}^{\|(1)}_V(x, P_z)\bigg\vert_{\mathrm{Fig}.~\ref{fig:feynman}(b)}=\frac{\alpha_s C_F}{2\pi}  \left\{\begin{aligned}
  &\bigg[\frac{x}{x_0(x-x_0)}\ln\frac{x}{x-x_0}-\frac{1}{2(x-x_0)}\bigg]_{+},~~~~&x<0\\
  &\bigg[\frac{x}{x_0(x-x_0)}\ln\frac{m^2_g}{4 P^2_z x_0(x_0-x)}+\frac{2x-x_0}{2x_0(x-x_0)}\bigg]_{+},~~~~&0<x<x_0\\
  &\bigg[\frac{x}{x_0(x-x_0)}\ln\frac{x-x_0}{x}+\frac{1}{2(x-x_0)}\bigg]_+,~~~~&x>x_0
  \end{aligned}
    \right.
\end{align}
and for Fig.~\ref{fig:feynman}(c), the results are
\begin{align}
  \phi^{\|(1)}_V(x, \mu)\bigg\vert_{\mathrm{Fig}.~\ref{fig:feynman}(c)}=\frac{\alpha_s C_F}{2\pi}\left\{\begin{aligned}
  &\bigg[\frac{x-1}{(x_0-1)(x-x_0)}\ln\frac{\mu^2 (1-x_0)}{m^2_g (1-x)}\bigg]_+,  &x_0<x<1\\
  &0, &\mathrm{others}\end{aligned}\right.
\end{align}
\begin{align}
  \t{\phi}^{\|(1)}_V(x,P_z)\bigg\vert_{\mathrm{Fig}.~\ref{fig:feynman}(c)}=\frac{\alpha_s C_F}{2\pi}\left\{\begin{aligned}
    &\bigg[\frac{x-1}{(x_0-1)(x-x_0)}\ln\frac{x-1}{x-x_0}-\frac{1}{2(x-x_0)}\bigg]_{+},~~~~&x<x_0\\
    &\bigg[-\frac{x-1}{(x_0-1)(x-x_0)}\ln\frac{m^2_g}{4P^2_z(1-x_0)(x-x_0)}\\
    &-\frac{2x-x_0-1}{2(x_0-1)(x-x_0)}\bigg]_+,~~~~&x_0<x<1\\
    &\bigg[\frac{x-1}{(x_0-1)(x-x_0)}\ln\frac{x-x_0}{x-1}+\frac{1}{2(x-x_0)}\bigg]_{+}.~~~~&x>1
  \end{aligned}\right.
\end{align}

For Fig.~\ref{fig:feynman}(d), we have
\begin{align}
  \phi^{\|(1)}_V(x, \mu)\bigg\vert_{\mathrm{Fig}.~\ref{fig:feynman}(d)}=0,
\end{align}
\begin{align}
  \t{\phi}^{\|(1)}_V(x, \mu)\bigg\vert_{\mathrm{Fig}.~\ref{fig:feynman}(d)}=\frac{\alpha_s C_F}{2\pi}\left\{\begin{aligned}
    &\bigg[\frac{1}{x-x_0}\bigg]_{+},~~~~&x<x_0\\
    &\bigg[-\frac{1}{x-x_0}\bigg]_+.~~~~&x>x_0
  \end{aligned}\right.
\end{align}

\subsection{Matching coefficients and evolution equations}
By using Eq.~\eqref{eq:match}, together with the one loop results under DR, one can determine the one loop corrections to the matching coefficients.
For $Z^{(1)}_{\perp}$, we have
\begin{align}\label{eq:match:trans:DR}
  Z^{(1)}_{\perp}(x,y,P_z,\mu)
  =C_F\left\{\begin{aligned}
    &\bigg[\frac{x}{y(x-y)}\ln\frac{x}{x-y}+\frac{x-1}{(y-1)(x-y)}\ln\frac{x-1}{x-y}\bigg]_+, &x<0<y\\
    &\bigg[\frac{x}{y(x-y)}\ln\frac{\mu^2}{4P^2_z x(y-x)}+\frac{x-1}{(y-1)(x-y)}\ln\frac{x-1}{x-y}+\frac{x}{y(x-y)}\bigg]_+, &0<x<y\\
    &\bigg[-\frac{x-1}{(y-1)(x-y)}\ln\frac{\mu^2}{4P^2_z(1-x)(x-y)}+\frac{x}{y(x-y)}\ln\frac{x-y}{x}\\
    &+\frac{1-x}{(y-1)(x-y)}\bigg]_+, &y<x<1\\
    &\bigg[\frac{x}{y(x-y)}\ln\frac{x-y}{x}+\frac{x-1}{(y-1)(x-y)}\ln\frac{x-y}{x-1}\bigg]_{+}, &y<1<x
  \end{aligned}\right.
\end{align}
and
for $Z^{(1)}_{\|}$, the result reads
\begin{align}\label{eq:match:long:DR}
  Z^{(1)}_{\|}(x,y,P_z,\mu)
  =C_F\left\{\begin{aligned}
    &\bigg[\frac{x-1}{y-1}\bigg(1+\frac{1}{x-y}\bigg)\ln\frac{x-1}{x-y}-\frac{x}{y}\bigg(1-\frac{1}{x-y}\bigg)\ln\frac{x}{x-y}\bigg]_{+}, &x<0<y\\
    &\bigg[\frac{x}{y}\bigg(1-\frac{1}{x-y}\bigg)\ln\frac{4P^2_z(y-x)x}{\mu^2}-\frac{x-1}{y-1}\bigg(1+\frac{1}{x-y}\bigg)\ln\frac{x-y}{x-1}\\
    &+\frac{x}{y(x-y)}+\frac{x}{y}\bigg]_+, &0<x<y\\
    &\bigg[\frac{x-1}{y-1}\bigg(1-\frac{1}{x-y}\bigg)\ln\frac{4P^2_z(1-x)(x-y)}{\mu^2}-\frac{x}{y}\bigg(1-\frac{1}{x-y}\bigg)\ln\frac{x-y}{x}\\
    &+\frac{1-x}{(y-1)(x-y)}+\frac{1-x}{1-y}\bigg]_+, &y<x<1\\
    &\bigg[\frac{x}{y}\bigg(1-\frac{1}{x-y}\bigg)\ln\frac{x}{x-y}-\frac{x-1}{y-1}\bigg(1+\frac{1}{x-y}\bigg)\ln\frac{x-1}{x-y}\bigg]_{+}. &y<1<x
  \end{aligned}\right.
\end{align}
In other regions, $Z^{(1)}_{\perp}$ and $Z^{(1)}_{\|}$ are zero.

Based on these matching coefficients one can also derive the $P_z$ evolution equations. Since the $\ln P_z$ dependence is the same in cut-off and DR schemes, the evolution equations are identical.

\section{Matching coefficients with a finite cut-off}\label{app:finitelambda}

In Sec.~\ref{sec:match}, we have calculated the matching coefficients under UV cut-off scheme. The cut-off $\Lambda$ has been taken to be $\Lambda\gg x P_z$. Since that it is difficult to
achieve the $\Lambda\to \infty$ limit for lattice simulations at present, $\Lambda$ and $x P_z$ could be of the same order. By considering the finite $\Lambda$ effect, the matching coefficients
presented by Eqs.~\eqref{eq:match:trans} and \eqref{eq:match:long} will be modified to be
\begin{align}
  Z^{(1)}_{\perp}(x, y, P_z, \Lambda)=&Z^{(1)}_{\perp}(x, y, P_z, \Lambda)\bigg\vert_{\mathrm{Eq.}\eqref{eq:match:trans}}+\delta Z^{(1)}_{\perp}(x, y, P_z, \Lambda),\\
    Z^{(1)}_{\|}(x, y, P_z, \Lambda)=&Z^{(1)}_{\|}(x, y, P_z, \Lambda)\bigg\vert_{\mathrm{Eq.}\eqref{eq:match:long}}+\delta Z^{(1)}_{\|}(x, y, P_z, \Lambda).
\end{align}
The corrections $\delta Z^{(1)}_{\perp}$ and $\delta Z^{(1)}_{\|}$ read
\begin{align}
  \delta Z^{(1)}_{\perp}(x, y, P_z, \Lambda)=&C_F\bigg[\frac{x}{y(x-y)}\bigg(\ln\frac{\Lambda(x)+P_z x}{\Lambda(x-y)+P_z(x-y)}+\frac{\Lambda(x-y)-\Lambda(x)}{2P_z}\bigg)+\frac{\Lambda(x-y)
  -\Lambda(0)}{2(x-y)^2 P_z}\bigg]_+\nonumber\\
  &+(x\to 1-x, ~y\to 1-y),\\
  \delta Z^{(1)}_{\|}(x, y, P_z, \Lambda)=&C_F\bigg[\frac{x}{y}\ln\frac{\Lambda(x)-P_z x}{\Lambda(x-y)-P_z(x-y)}+\frac{x}{y(x-y)}\bigg(
  \ln\frac{\Lambda(x)+P_z x}{\Lambda(x-y)+P_z(x-y)}+\frac{\Lambda(x-y)-\Lambda(x)}{2P_z}\bigg)\nonumber\\
  &+\frac{\Lambda(x-y)-\Lambda(0)}{2(x-y)^2 P_z}\bigg]_+\nonumber\\
  &+(x\to 1-x, ~y\to 1-y),
\end{align}
where $\Lambda(x)\equiv \sqrt{\Lambda^2+x^2 P^2_z}$. One can examine that $\delta Z^{(1)}_{\Gamma}\to 0$ when $\Lambda\to \infty$.

\end{document}